%% file: gravity_essay_17.tex
\documentclass[11pt]{article}

\usepackage{authblk}
\usepackage{amsmath,amssymb}
\usepackage{hyperref}
\usepackage{graphicx}
\hypersetup{
    colorlinks,
    citecolor=black,
    filecolor=black,
    linkcolor=black,
    urlcolor=black
}
\usepackage[super,compress]{cite}

\input{newcommands}

\begin{document}

\markboth{Robert J. Hardwick, Vincent Vennin and David Wands}
{A window onto early inflation}

\title{A Quantum Window Onto Early Inflation}

\author{Robert J. Hardwick, Vincent Vennin, David Wands}

\affil{Institute of Cosmology \& Gravitation, \\
University of Portsmouth, Dennis Sciama Building, \\
Burnaby Road, Portsmouth, PO1 3FX, United Kingdom\\
Corresponding author: robert.hardwick@port.ac.uk \\
\emph{Keywords:} Inflation, Physics of the Early Universe, Quantum Gravity Phenomenology \\
\emph{PACS Number(s):} 98.80.Cq 04.60.Bc 04.62.+v 05.10.Gg
}

\maketitle

\begin{abstract}
Inflation in the early Universe is one of the most promising probes of gravity in the high-energy regime. However, observable scales give access to a limited window in the inflationary dynamics. In this essay, we argue that quantum corrections to the classical dynamics of cosmological fields allow us to probe much earlier epochs of the inflationary phase and extend this window by many orders of magnitude. We point out that both the statistics of cosmological fluctuations at observable scales, and the field displacements acquired by spectator fields that play an important role in many post-inflationary processes, are sensitive to a much longer phase of the inflationary epoch.
\end{abstract}

\begin{center}
\textit{Essay written for the Gravity Research Foundation 2017 Awards for Essays on Gravitation.}
\end{center}

\vspace{0.5cm}

Cosmological inflation~\cite{Starobinsky:1980te, Sato:1980yn, Guth:1980zm, Linde:1981mu, Albrecht:1982wi, Linde:1983gd, Olive:2016xmw} is one of the only places in physics where an effect based on general relativity (the accelerated expansion) and quantum mechanics (the amplification of vacuum quantum fluctuations of the gravitational and matter fields to large-scale cosmological perturbations~\cite{Starobinsky:1979ty, Mukhanov:1981xt, Hawking:1982cz,  Starobinsky:1982ee, Guth:1982ec, Bardeen:1983qw}) gives rise to predictions that can be tested experimentally. For this reason, it is an ideal framework to discuss fundamental issues related to the nature of gravity and even quantum gravity. Indeed, the cosmological perturbations produced in this early epoch have been measured to very high accuracy~\cite{Adam:2015rua, Ade:2015lrj} in the temperature and polarisation anisotropies of the cosmic microwave background (CMB), and can be probed more generally in the large-scale structure of our Universe.

At a given physical length scale $ a / k$, where $a$ is the scale factor of the Universe and $k$ is a fixed comoving wavenumber, the statistical properties of cosmological fluctuations are mostly determined by the properties of the inflationary classical dynamics around the time when $a/k$ crosses the Hubble radius $H^{-1}\equiv a/\dot{a}$, where a dot denotes differentiation with respect to cosmic time $t$. For example, if inflation is driven by a single scalar field $\phi$ with potential $V(\phi)\equiv 24 \pi^2\Mp^4 v(\phi)$ (where $\Mp$ is the Planck mass), the power spectrum of curvature perturbations $\zeta$ at scale $k$ is given by
\bea
\label{eq:Pzeta:class}
\calP_\zeta(k) = \frac{2}{\Mp^2}\frac{v^3[\phi_*(k)]}{{v^\prime}^2[\phi_*(k)]},
\eea
where $\phi_*(k)$ is the value of $\phi$ when $a/k$ exits the Hubble radius. The range of scales probed \eg in the CMB then translates into a time interval during inflation of length $N\sim 7$, measured by the number of \efolds $N\equiv \ln(a)$. If one includes the large-scale structure of our Universe, this window is extended but cannot exceed the last $\sim 60$ \efolds of inflation. But can we ever learn about larger scales, hence earlier times?

During inflation, the coarse-grained fields (\ie scales larger than the Hubble radius) are constantly sourced by the small-wavelength quantum fluctuations as they cross the Hubble radius. This quantum backreaction on the dynamics of the Universe can be modeled through the stochastic inflation formalism~\cite{Starobinsky:1986fx}, where $\phi$ does not only evolve classically under the gradient of its potential $V(\phi)$, but also receives quantum corrections through an additional noise term
\bea
\label{eq:Langevin}
\frac{\dd \phi}{\dd N} = -\frac{V^\prime(\phi)}{3H^2}+\frac{H}{2\pi}\xi ,
\eea
where $\xi$ is a normalised white Gaussian noise. The system then explores parts of the potential that would be inaccessible under the classical dynamics. For example, the power spectrum~(\ref{eq:Pzeta:class}) is now given by~\cite{Vennin:2015hra, Assadullahi:2016gkk, Vennin:2016wnk}
\bea
\label{eq:Pzeta:sto}
\calP_\zeta(k) = &2
\left\lbrace \int_{\phi_*}^{\infty}\frac{\mathrm{d} x}{M_{{}_\mathrm{Pl}}}\frac{1}{v\left(x\right)}\exp\left[\frac{1}{v\left(x\right)}-\frac{1}{v\left(\phi_*\right)}\right] \right\rbrace^{-1}
\times \\
&
\int_{\phi_*}^{\infty}\frac{\mathrm{d} x}{M_{{}_\mathrm{Pl}}}\left\lbrace\int_{x}^{\infty} \frac{\mathrm{d} y}{M_{{}_\mathrm{Pl}}} \frac{1}{v\left(y\right)}\exp\left[\frac{1}{v\left(y\right)}-\frac{1}{v\left(x\right)}\right] \right\rbrace^2.
\eea
Contrary to \Eq{eq:Pzeta:class}, this expression does not only depend on the potential evaluated at $\phi_*(k)$, but relies on the properties of the potential in the entire inflationary domain. For this reason, even the limited range of scales probed in the CMB may contain imprints from early features of the inflationary dynamics and in this sense, quantum diffusion in an expanding background greatly extends the observational window. In practice, when $v\ll 1$, \ie $V\ll \Mp^4$, \Eq{eq:Pzeta:sto} is well approximated by \Eq{eq:Pzeta:class} so the dependence on the potential function outside the standard observational window is usually Planck suppressed. This is however not the case when several fields drive inflation~\cite{Kawasaki:2015ppx, Assadullahi:2016gkk, Vennin:2016wnk}, or in very flat regions of the potential that can drive the dynamics at smaller (but still accessible~\cite{Pattison:2017mbe}) scales than the ones probed in the CMB.

\begin{figure}[h!]
\begin{center}
\includegraphics[width=0.6\textwidth]{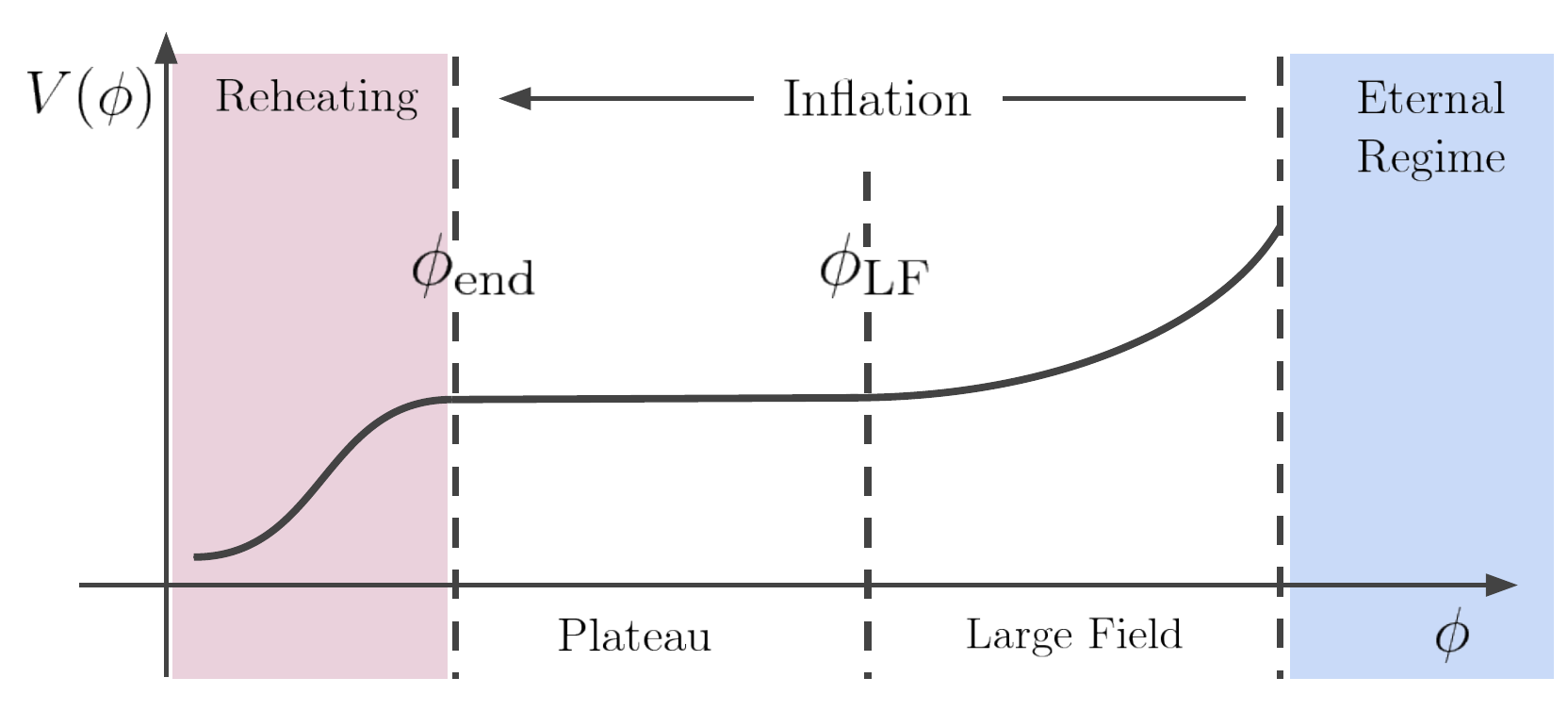}
\caption{\label{fig:potential} Toy inflationary potential considered in this work, made of a plateau (\ie asymptotically constant) part between $\phi_\uend$ and $\phi_{\mathrm{LF}}$, and a monomial large-field part (where $V \propto \phi^p$) at $\phi>\phi_{\mathrm{LF}}$. CMB observations constrain the number of \efolds spent on the plateau to be $N_{\mathrm{plateau}}>60$, while the dynamics of spectator fields is sensitive on a much wider part of the inflationary potential.}
\end{center}
\end{figure}
Another, less direct but more sensitive, cosmological probe sensitive to the early stages of inflation through quantum diffusion is the field displacement acquired by light spectator fields~\cite{Enqvist:2012xn, Sanchez:2016lfw, Hardwick:2017fjo}. Such fields do not participate in the accelerated expansion of inflation but can play an important role afterwards (\eg Higgs fields~\cite{Figueroa:2016dsc} or dark matter candidates~\cite{Nurmi:2015ema}) that depends on the field displacement they acquire during inflation. Let us consider the toy model depicted in \Fig{fig:potential} where the inflaton potential $V(\phi)$ is made of a plateau (\ie asymptotically constant) part between $\phi_\uend$ and $\phi_{\mathrm{LF}}$ and a monomial large-field (\ie $V\propto \phi^p$) part at $\phi>\phi_{\mathrm{LF}}$. Observations of the CMB constrain the potential to be of the plateau type in the last few \efolds of inflation~\cite{Martin:2013nzq} so in the standard setup, the only constraint one has is that $\phi_{\mathrm{LF}}$ should be located at least $\sim 60$ \efolds before the end of inflation.

\begin{figure}[h!]
\begin{center}
\includegraphics[width=0.48\textwidth]{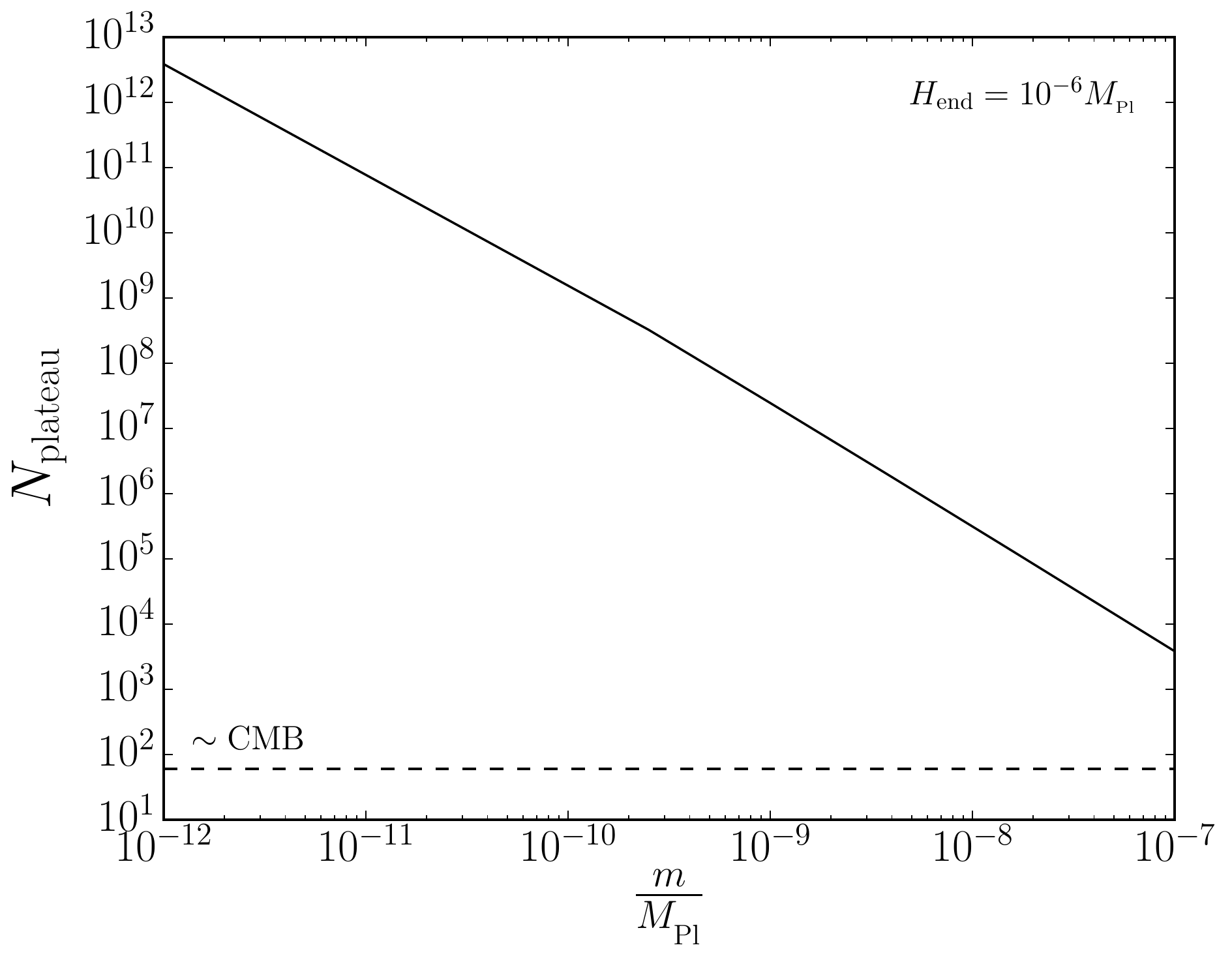}
\includegraphics[width=0.48\textwidth]{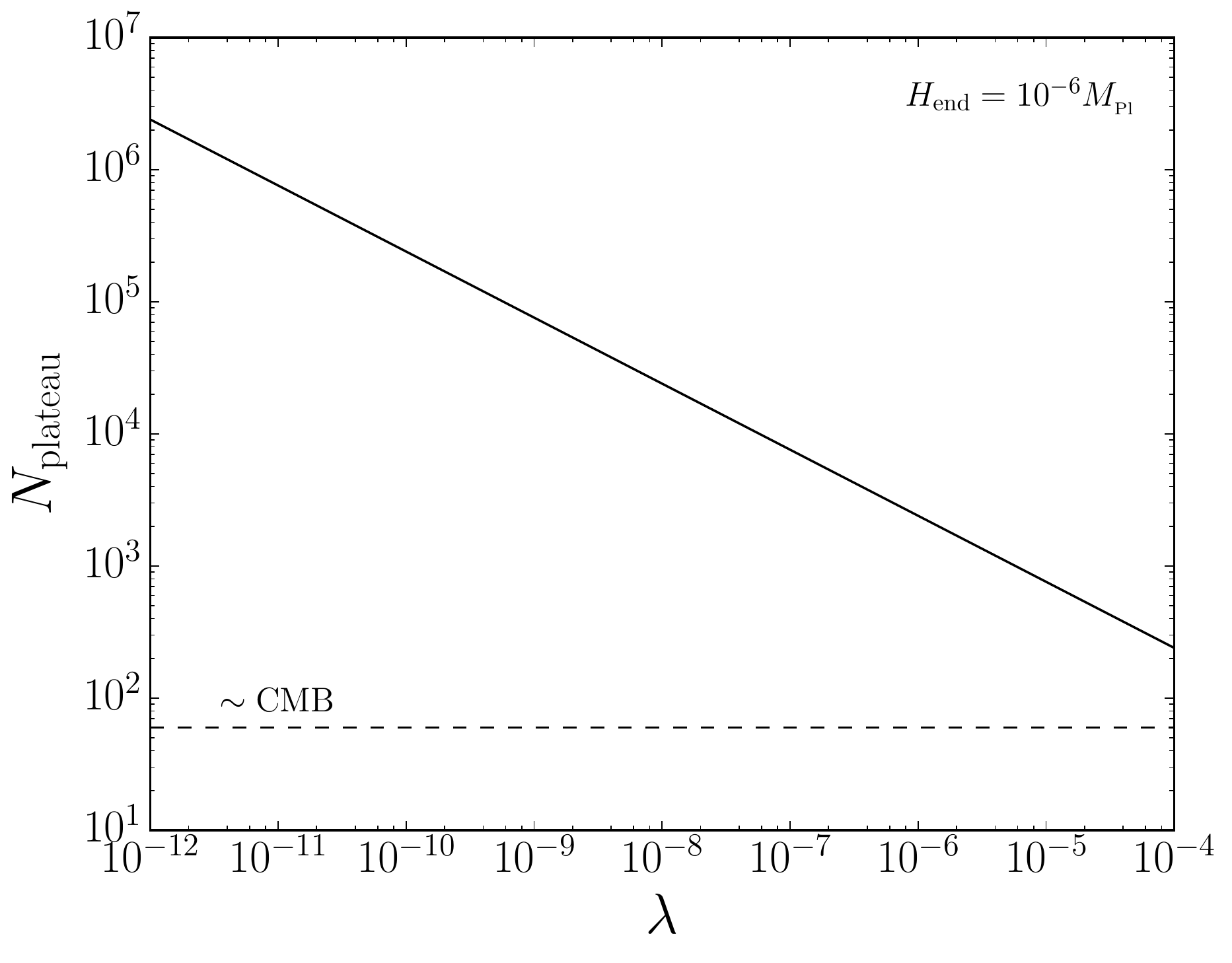}
\caption{\label{fig:deltaN} Minimum number of \efolds spent on the plateau part of the inflationary potential so that the spectator field displacement at the end of inflation is independent of the large-field correction to the inflaton potential. The left panel corresponds to a quadratic spectator, $V(\sigma) = m^2\sigma^2/2$ and the right panel corresponds to a quartic spectator, $V(\sigma)=\lambda \sigma^4$. Through CMB observations interpreted in the standard way, one gets the constraint $N_{\mathrm{plateau}}>60$ (denoted with the dashed line), while spectator fields are sensitive to a much wider part of the inflationary dynamics.}
\end{center}
\end{figure}
A spectator field $\sigma$ on top of this inflationary background evolves under its potential $V(\sigma)$ and according to \Eq{eq:Langevin} (where $\phi$ is to be replaced by $\sigma$). If $H$ is constant, the probability distribution $\Pr (\sigma )$ relaxes towards the stationary attractor profile~\cite{Starobinsky:1986fx}
\bea \label{eq:Pstat}
\Pr (\sigma ) \propto \exp \left[ -\frac{8\pi^2 V (\sigma )}{3H^4}\right]
\eea
where any initial condition is erased. However, since $H$ is not exactly constant during inflation, this does not always happen on the large-field part of the inflationary potential, since the relaxation time towards \Eq{eq:Pstat} can be larger than the variation time scale of $H$ there. For example, if the spectator potential is quadratic, $V(\sigma) = m^2 \sigma^2 /2 $, \Eq{eq:Pstat} can never be attained in the early phase of large-field evolution where the typical field displacement remains strongly dependent on initial conditions. By setting $\sigma=0$ at the exit point of eternal inflation (where the dynamics of $\phi$ is itself dominated by stochastic corrections), one can derive a lower bound on the number of \efolds $N_{\mathrm{plateau}}$ spent on the plateau part of the inflaton potential so that the details of the large-field phase are erased from the distribution of $\sigma$ at the end of inflation~\cite{Hardwick:2017fjo},
\bea
N_{\mathrm{plateau}} \geq \frac{3H_{\mathrm{plateau}}^2}{2m^2} \ln \left[ \frac{8\pi p \, m^2 \Mp^2 }{3H_{\mathrm{plateau}}^4 (p+2)} \right]
\eea
for $p\geq 2$. It is displayed in the left panel of \Fig{fig:deltaN} for $p=2$ (but the result depends only mildly on $p$). Compared to the standard constraint $N_{\mathrm{plateau}} \geq 60$, one can see that the observational window on the inflaton potential extends by orders of magnitude. For a quartic spectator $V(\sigma) = \lambda \sigma^4$, it turns out that \Eq{eq:Pstat} is adiabatically tracked at early time in the large-field phase (more precisely, when $H>\lambda^{-p/8} H_\mathrm{plateau}$). In this case, initial conditions on the spectator field displacement can be erased during this adiabatic epoch, and the minimal number of \efolds spent on the plateau such that no imprint is left from the large-field epoch on the distribution of $\sigma$ at the end of inflation is given by~\cite{Hardwick:2017fjo}
\bea
\label{eq:quartN}
N_{\mathrm{plateau}} \geq \frac{4\pi \Gamma \left( \frac{3}{4}\right)}{\Gamma \left( \frac{1}{4}\right)} \sqrt{\frac{2}{3\lambda}} \ln (2) ,
\eea
for $p \geq 2$. It is displayed in the right panel of \Fig{fig:deltaN} where one can see again that the observational window on the inflaton potential extends by orders of magnitude.

Thus the quantum dynamics of cosmological fields in the early Universe gives access to a vast range of scales that extend the classical window by orders of magnitude and allow us to explore high-energy gravity beyond the observable horizon.

\section*{Acknowledgments}
We are grateful to our collaborators Chris Byrnes,  Alexei Starobinsky and Jes\'us Torrado.
This work was supported by STFC grants ST/N000668/1, ST/K502248/1 and ST/N504245/1.

\bibliography{gravity_essay_17}
\bibliographystyle{JHEP}

\end{document}

%% file: newcommands.tex
\newcommand{\ie}{{i.e.~}}

\newcommand{\eg}{\textsl{e.g.~}}

\newcommand{\dd}{\mathrm{d}}

\newcommand{\sss}[1]{{\scriptscriptstyle{#1}}}

\newcommand{\uPl}{\mathrm{Pl}}

\newcommand{\uend}{\mathrm{end}}

\newcommand{\usssPl}{\sss{\uPl}}

\newcommand{\calP}{\mathcal{P}}

\newcommand{\Mp}{M_\usssPl}

\newcommand{\efolds}{$e$-folds~}

\newcommand{\beq}{\begin{equation}}
\newcommand{\eeq}{\end{equation}}
\newcommand{\bea}{\begin{equation}\begin{aligned}}
\newcommand{\eea}{\end{aligned}\end{equation}}

\newlength{\wsingfig}
\newlength{\wdblefig}
\newlength{\wquadfig}
\newlength{\wtriplefig}
\newcommand{\Eq}[1]{Eq.~(\ref{#1})}

\newcommand{\Fig}[1]{Fig.~{\ref{#1}}}